\documentclass[preprint,aps,superscriptaddress,]{revtex4}

\usepackage{graphicx}
\usepackage{dcolumn}
\usepackage{bm}
\usepackage{eucal}
\usepackage{color}
\usepackage{multirow}
\usepackage{citesort}

\def\bfa{BaFe$_2$As$_{2}$}
\def\sfa{SrFe$_2$As$_{2}$}
\def\tc{$T_{\rm sc}$}
\def\tn{$T_{\rm N}$}
\def\tz{$T_0$}
\def\mub{${\mu}_{\rm B}$}

\begin{document}

\title{Columnar magnetic structure coupled with orthorhombic distortion in the Antiferromagnetic Iron Arsenide SrFe$_2$As$_2$}

\author{K. Kaneko}
\email{kaneko.koji@gmail.com}
\affiliation{Max-Planck-Institut f\"ur Chemische Physik fester Stoffe, N\"othnitzer Str. 40, 01187 Dresden, Germany }
\affiliation{Advanced Science Research Center, Japan Atomic Energy Agency, Tokai, Naka, Ibaraki 319-1195, Japan}

\author{A. Hoser}
\affiliation{ Helmholtz-Zentrum Berlin f\"ur Materialien und Energie, Glienicker Str. 100, 14109 Berlin, Germany}

\author{N. Caroca-Canales}
\affiliation{Max-Planck-Institut f\"ur Chemische Physik fester Stoffe, N\"othnitzer Str. 40, 01187 Dresden, Germany }

\author{A. Jesche}
\affiliation{Max-Planck-Institut f\"ur Chemische Physik fester Stoffe, N\"othnitzer Str. 40, 01187 Dresden, Germany }

\author{C. Krellner}
\affiliation{Max-Planck-Institut f\"ur Chemische Physik fester Stoffe, N\"othnitzer Str. 40, 01187 Dresden, Germany }

\author{O. Stockert}
\affiliation{Max-Planck-Institut f\"ur Chemische Physik fester Stoffe, N\"othnitzer Str. 40, 01187 Dresden, Germany }

\author{C. Geibel}
\affiliation{Max-Planck-Institut f\"ur Chemische Physik fester Stoffe, N\"othnitzer Str. 40, 01187 Dresden, Germany }

\date{\today}

\begin{abstract}
Neutron powder diffraction experiments were carried out on polycrystalline SrFe$_2$As$_2$ in order to determine the magnetic structure and its relationship with the crystallographic one.
Below $T_0$=205\,K, magnetic reflections appear simultaneously with the onset of the orthorhombic distortion.
From a detailed Rietveld analysis, the magnetic propagation vector of {\sfa} is determined to be {\textit{\textbf{q}}}=(1\,0\,1); 
the coupling of Fe moments is antiferromagnetic along the longer $a$ direction within the Fe-As layer, and the interlayer coupling is antiferromagnetic as well.
The size of the Fe magnetic moment is determined to be 1.01(3)\,${\mu}_{\rm B}$ with an orientation parallel to the $a$-axis.
The temperature dependence of the magnetic moment shows excellent agreement with not only that of the muon precession frequency but also with that of the structural distortion,  
revealing the strong coupling of the columnar magnetic order and the structural distortion in {\sfa}. 
\end{abstract}

\maketitle

The discovery of superconductivity in doped LaFeAsO triggered intensive research on layered FeAs systems\cite{LaFeAsOF_YKamihara_01}.
So far, the superconducting (SC) transition temperature {\tc} of $R$FeAsO has rapidly been raised up to 54\,K\cite{PrFeAsOF_ZARen_01,NdFeAsOF_ZARen_01, CeFeAsOF_GFChen_01,SmOFFeAs_XHChen_01,GdFeAsO_JYang_01}.
Furthermore, a similar high {\tc} exceeding 35\,K was also discovered upon doping the related compounds $A$Fe$_2$As$_2$ ($A$=Ca, Sr, Ba).\cite{AFe2As2_CKrellner_01,SrFe2As2_GFChen_01}
$A$Fe$_2$As$_2$ has the well-known ThCr$_2$Si$_2$-type structure, which can be regarded as replacing the (R$_2$O$_2$)$^{2+}$ layer in $R$FeAsO by a single divalent ion ($A^{2+}$) layer keeping the same electron count.\cite{AFe2As2_CKrellner_01}
$R$FeAsO and $A$Fe$_2$As$_2$ were found to present very similar physical properties.
Undoped $R$FeAsO compounds ($R$=La-Gd) present a structural transition at $T_0{\sim}$150\,K followed by the formation of a spin-density wave (SDW) at a slightly lower temperature {\tn}${\sim}$140\,K.
$A$Fe$_2$As$_2$ exhibits a similar structural distortion, whereas the SDW forms at the same or a slightly lower temperature\cite{SrFe2As2_MTegel_01}.
Electron or hole doping leads to the suppression of the SDW and to the onset of superconductivity.
This connection between a vanishing magnetic transition and the simultaneous formation of a SC state is reminiscent of the behavior in the cuprates and in the heavy-fermion systems, suggesting the SC state in these doped FeAs systems to be of unconventional nature. 
In addition, the SDW state is strongly coupled to the lattice distortion in this system and therefore it is of high interest to reveal the relationship between the lattice, the magnetism and superconductivity.

The magnetic structures of the FeAs system have been investigated by neutron diffraction study on $R$FeAs(O,F) for $R$=La, Ce, and Nd, and {\bfa}\cite{LaFeAsOF_CdlCruz_01,CeFeAsOF_JZhao_01,BaFe2As2_QHuang_01}.
The antiferromagnetic (AFM) reflection was observed around the (1,0) or (0,1) position with respect to the Fe-As layer.
This corresponds to Fe moments with AFM coupling along the $a$ or $b$ direction, called the columnar structure.
In contrast, interlayer couplings is not unique among these compounds, suggesting a relatively weak coupling between the layers.
The size of the estimated Fe moment is  less than 1\,{\mub}, which is much smaller than theoretical predictions.
So far the magnetic structure of these iron arsenides have not been determined uniquely, especially the relation of the direction of the AFM ordering with respect to the short or long Fe-Fe distances is not settled.
This is one of the fundamental questions on the magnetic properties of these materials and is indispensable for understanding the interplay between magnetism and superconductivity.

Among the reported FeAs systems, {\sfa} should be a suitable compound to study the detailed magnetic structure.
By the substitution of Sr with K or Cs, the superconductivity appears with the maximum {\tc} of  38\,K.\cite{SrFe2As2_GFChen_01,SrFe2As2_KSasmal_01}
The parent compound {\sfa} undergoes a first order transition at $T_0$=205\,K, where both, the SDW and the structural transition, occur simultaneously.\cite{SrFe2As2_AJesche_01}
A detailed x-ray diffraction study clarified that the structural transition at {\tz} is from tetragonal($I4/mmm$) to orthorhombic($Fmmm$) and therefore similar to BaFe$_2$As$_2$\cite{SrFe2As2_MTegel_01,SrFe2As2_AJesche_01}.
A stronger magnetism in {\sfa} is inferred from the higher ordering temperature, the larger value of the Pauli like susceptibility above $T_0$ as well as the larger Fe hyperfine field observed in M\"ossbauer experiments.\cite{SrFe2As2_MTegel_01,SrFe2As2_AJesche_01}
Therefore, a larger ordered moment is expected for this compound, which should allow a detailed analysis.

In this paper, we report a neutron powder diffraction study on {\sfa}.
The precise analysis of the observed magnetic Bragg peaks on {\sfa}\,\,allowed us to uniquely determine the magnetic structure.
The magnetic propagation vector of {\sfa} is {\textit{\textbf{q}}}=(1\,0\,1), thus the AFM coupling is realized in the longer Fe-Fe direction within the Fe-As layer.
The stacking along the $c$-direction is also AFM.
The direction of the magnetic moment is parallel to the $a$-axis as well. 
A remarkable agreement was obtained in the temperature evolution of the magnetic moment and structural distortion obtained from independent measurements.
These facts clearly demonstrate the close relationship between magnetism and lattice distortion in {\sfa}.

The details on the sample preparation of {\sfa} have been described in Ref.~\onlinecite{AFe2As2_CKrellner_01}.
Neutron powder diffraction experiments were carried out on the two-axis diffractometer E6, installed at Helmholtz Center Berlin, Germany.
A double focusing pyrolytic graphite (PG) monochromator results in a high neutron flux at the sample position.
Data were recorded at scattering angle up to 110$^{\circ}$ using a two dimensional position sensitive detector (2-D PSD) with the size of 300${\times}$300\,mm$^2$.
Simultaneous use of the double focusing monochromator and 2-D PSD in combination with the radial oscillating collimator gives a high efficiency for taking diffraction patterns.
The neutron wavelength was chosen to be 2.4\,\AA\,\,in connection with a PG filter in order to avoid higher-order contaminations.
As a compensation for the high efficiency, error in the absolute values in  the lattice constants become relatively large.
Since the detailed absolute values were already obtained from the x-ray diffraction\cite{SrFe2As2_AJesche_01}, we relied on these data and focus here on the details of the magnetic structure.
Fine powder of {\sfa} with a total mass of ${\sim}$2\,g was sealed in a vanadium cylinder as sample container.
The standard $^4$He cryostat was used to cool the sample down to 1.5\,K well below {\tz}.
Neutron diffraction patterns were taken at different temperatures between 1.5\,K and 220\,K and the obtained diffraction patterns were analyzed by the Rietveld method using the software RIETAN-FP\cite{RIETAN_FIzumi_01}.
The software VESTA\cite{VESTA_KMomma_01} was used to draw both crystal and magnetic structures.

\begin{figure}[t]
	\begin{center}
		\includegraphics[width=10cm]{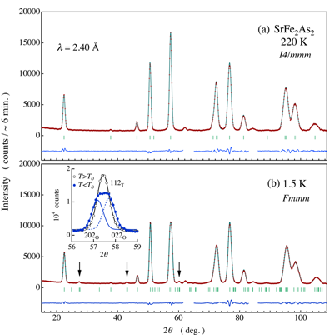}
	\end{center}
	\caption{Neutron powder diffraction patterns measured on E6 at (a)\,220\,K ($T>T_0$) and (b)\,1.5\,K ($T<T_0$). The result of the Rietveld analysis, residual curves and the calculated peak positions indicated by tick marks are included in each panel. The inset shows the temperature variation of the peak profile around 2${\theta}$=57.5$^{\circ}$. }
	\label{f1}
\end{figure}

Figure~\ref{f1} shows neutron diffraction patterns of {\sfa} taken at (a)\,$T$=220\,K ($T>T_0$) and  (b)\,$T$=1.5\,K ($T<T_0$).
Results of the Rietveld analysis, the residual intensity curve, and tick marks indicating the expected reflection angle, are also plotted in the figure.
For the high temperature $T>${\tz}, the observed pattern is well reproduced by assuming the crystal structure with the space group $I4/mmm$ as shown in Fig.~\ref{f1}(a).
These results are consistent with that reported from x-ray diffraction study, including the positional parameter of As, $z_{\rm As}$=0.3602(2), where the number in the parenthesis indicate the uncertainty at the last decimal point.
The conventional reliability factors $R_{\rm wp}$=3.73\,\% $R_{\rm I}$=3.18\,\% and $R_{\rm F}$=1.93\,\% of the present analysis indicate the high quality of the present analysis.
Small impurity peaks observed at around 63$^{\circ}$ and 84$^{\circ}$ which originate from Cu and Al were excluded from the analysis.

Below {\tz}=205\,K, some nuclear Bragg reflections became broad.
The inset of Fig.~\ref{f1} shows the Bragg peak profile around 57.5$^{\circ}$ where the 1\,1\,2 Bragg peak of the high temperature tetragonal phase was observed.
The peak intensity drops and broadens on passing through {\tz}, as expected for the orthorhombic distortion where the lattice constant $a$ becomes longer than $b$.
The reflection profile could be well reproduced by the lattice distortion from the tetragonal to orthorhombic lattice reported by the x-ray diffraction study\cite{SrFe2As2_AJesche_01}.
Within the errors, the positional parameter of As in the orthorhombic phase $z_{\rm As}$=0.3604(2) is the same as that above $T_0$.

\begin{figure}[t]
	\begin{center}
		\includegraphics[width=10cm]{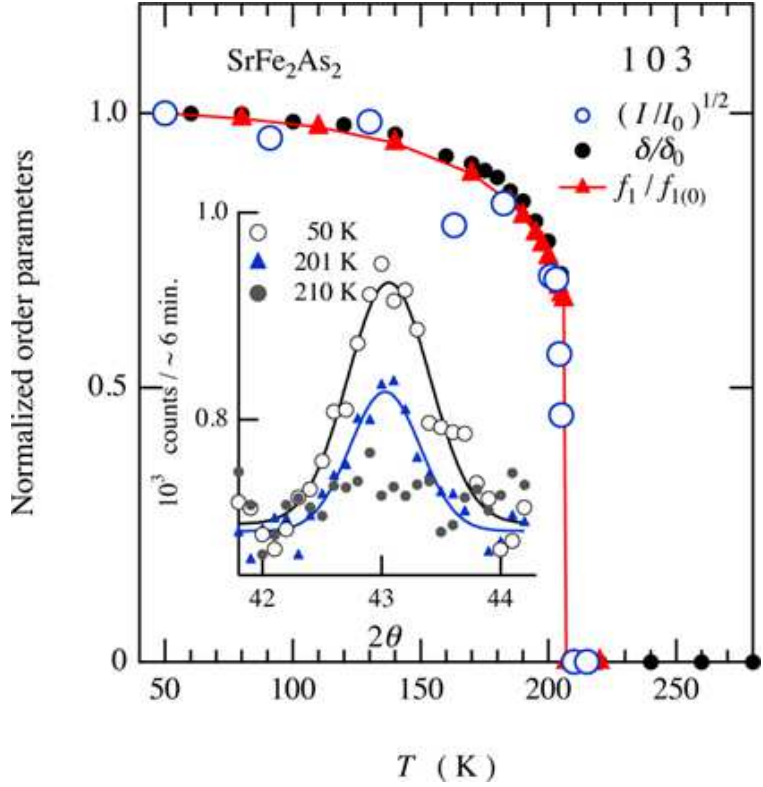}
	\end{center}
	\caption{Temperature dependence of the 1\,0\,3 magnetic reflection intensity normalized to that at 1.5\,K, plotted as ${\sqrt{I/I_0}}$, which is proportional to the size of the ordered moment.
	Closed circles and triangles correspond to the temperature dependence of normalized lattice distortion and the muon precession frequency $f_1$, respectively,  of {\sfa} taken from Ref. \onlinecite{SrFe2As2_AJesche_01}. The inset shows the 1\,0\,3 magnetic peak profile taken at 50, 201 and 210\,K.}
	\label{f2}
\end{figure}

In addition to the lattice distortion, additional reflections were also observed at $T$=1.5\,K as indicated by arrows in Fig.~\ref{f1}(b).
These superlattice peaks are most prominent at low scattering angles, and the corresponding peaks were not observed in the x-ray diffraction\cite{SrFe2As2_MTegel_01,SrFe2As2_AJesche_01}.
Therefore, the origin of these superlattice peaks should be magnetic.
Figure~\ref{f2} shows the temperature dependence of the integrated intensity of the superlattice reflection around 43$^{\circ}$ which can be indexed as 1\,0\,3 as described later.
The left axis is set to be proportional to the size of the magnetic moment, ${\sqrt{I/I_0}}$, where $I_0$ is the intensity at the lowest temperature.
The 1\,0\,3 reflection appears below $T_0$= 205\,K and shows a sharp increase in its intensity, which is clearly seen in the profile shown in the inset.
The magnetic moment at 201\,K, just 4\,K below {\tz}, already attains 70\,\% of that at 1.5\,K.
These findings strongly support the first order transition at {\tz} in {\sfa}.
In the same figure, the relative lattice distortion and the muon precession frequency of {\sfa} taken from Ref.\onlinecite{SrFe2As2_AJesche_01} both normalized to the saturated values are plotted.
A remarkable agreement of the temperature dependences is seen for these quantities obtained from independent measurements.
This clearly demonstrates the close relationship between magnetism and lattice distortion in {\sfa}.

\begin{figure}[t]
	\begin{center}
		\includegraphics[width=10cm]{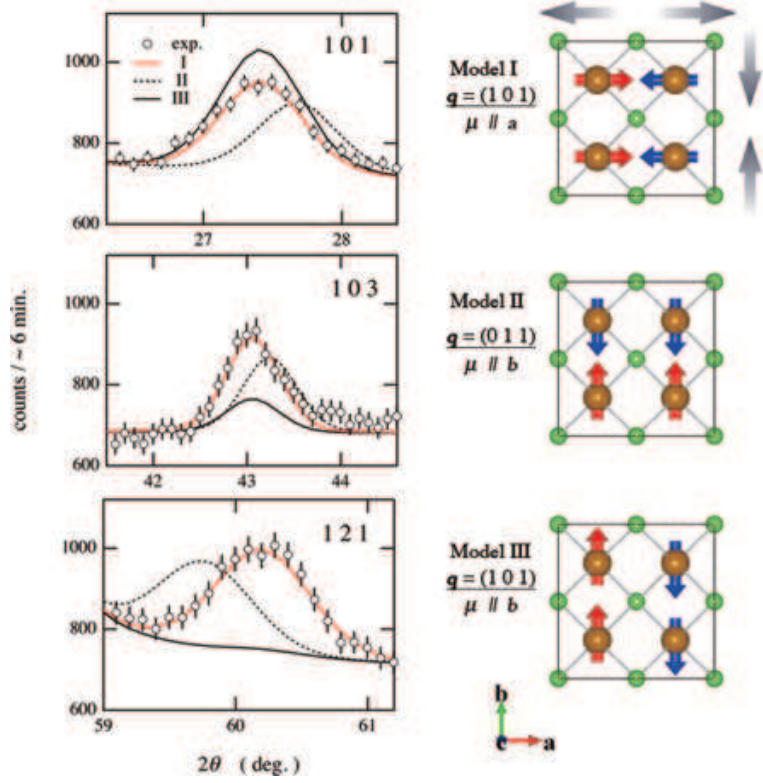}
	\end{center}
	\caption{Magnetic Bragg peak profiles at (1\,0\,1), (1\,0\,3) and (1\,2\,1).
	Red(gray), dotted and solid lines are the calculated intensity assuming the respective magnetic structure model:
	(i)\,{\textit{\textbf{q}}}=(1\,0\,1), ${\mu}{\parallel}a$, (ii)\,{\textit{\textbf{q}}}=(0\,1\,1), ${\mu}{\parallel}b$, (iii)\,{\textit{\textbf{q}}}=(1\,0\,1), ${\mu}{\parallel}b$.
	Shaded arrows illustrate the orthorhombic distortion concomitantly occurs at $T_0$.}
	\label{f3}
\end{figure}

Hereafter, we analyze the magnetic structure of {\sfa}.
For simplicity, in the magnetic structure analysis the structural parameters were fixed to the best value determined from the nuclear Bragg peaks. 
In the following, three representative models shown in the right panel of Fig.~\ref{f3} are considered.
At first, the direction of the AFM coupling with respect to the orthogonal axis is examined.
This corresponds to the difference between Model I with the propagation vector {\textbf{\textit{q}}}=(1\,0\,1) and Model II with {\textbf{\textit{q}}}=(0\,1\,1).
The difference between these models appears in the scattering angle arising from the subtle difference between $a$ and $b$.
The difference  can be clearly seen in the comparison for 1\,0\,1(upper panel) and 1\,2\,1(bottom) reflections.
When Model I is used as a reference, the Model II gives a higher scattering angle for 1\,0\,1 which then corresponds to 0\,1\,1, and a lower angle for 1\,2\,1 (2\,1\,1).
The comparison in Fig.~\ref{f3} clearly show that only Model I gives the correct positions of the 1\,0\,1 and 1\,2\,1 reflections.
Therefore, the magnetic propagation vectors of {\sfa} is determined to be {\textbf{\textit{q}}}=(1\,0\,1).

\begin{figure}[t]
	\begin{center}
		\includegraphics[width=9cm]{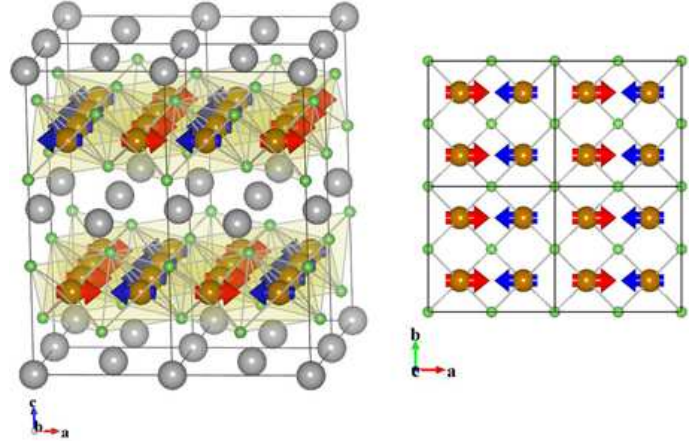}
	\end{center}
	\caption{Magnetic and crystal structure of {\sfa} in the ground state.
	The AFM coupling is along the $a$ direction having the longer Fe-Fe distance, corresponding to {\textit{\textbf{q}}}=(1\,0\,1).
	The Fe magnetic moment lies within the Fe plane and is aligned parallel to the $a$ axis.}
	\label{f4}
\end{figure}

In a second step, we try to analyze the direction of AFM moment, which corresponds to Model I and III.
In Model I, the magnetic moment is set to be parallel to the AFM coupling in the Fe-As plane, ${\mu}{\parallel}a$, whereas it is perpendicular in Model III.
The magnetic diffraction intensity is depending on the angle between the magnetic moment and scattering vector.
The powder average of this angle factor for the orthorhombic symmetry is given as follows;
\begin{equation}
{\langle}q^2{\rangle}=1-(h^2a^{*2}\cos^2{\psi}_a+k^2b^{*2}\cos^2{\psi}_b+l^2c^{*2}\cos^2{\psi}_c)d^2
\end{equation}
where $h, k, l$ are the reflection indexes, $a^*, b^*, c^*$ are the primitive lattice vectors in reciprocal space, and $d$ is the spacing of the ($hkl$) in real space.
${\psi}_a$, ${\psi}_b$, ${\psi}_c$ are the angles between the magnetic moment and crystallographic axes $a, b$ and $c$. 
This factor affects the relative magnetic intensities of 1\,0\,1, 1\,0\,3, and 1\,2\,1 diffraction peaks.
Model I gives the comparable intensity for all three peaks, in good agreement with experimental results.
In contrary, Model III results in a strong 1\,0\,1 and a vanishing 1\,2\,1 peak, which does not at all fit with the measured intensities.
As a result, the calculation based on the Model I gives an excellent agreement for the position and intensity of all magnetic reflections.
Therefore, we can conclusively determine the direction of the magnetic moment to be $a$.
By using the structure Model I, the size of the Fe magnetic moment is deduced to be 1.01(3)\,{\mub} with good reliability factors $R_{\rm wp}$=3.54\,\% $R_{\rm I}$=2.49\,\% and $R_{\rm F}$=1.63\,\%.
The determined magnetic structure of {\sfa} is shown in Fig.~\ref{f4}; the AFM coupling occurs along the longer $a$ direction and the moment orients in the same direction.
The magnetic ordering in {\sfa} within the FeAs layer seems identical to that in \bfa\cite{BaFe2As2_QHuang_01}, in LaFeAs(O,F)\cite{LaFeAsOF_CdlCruz_01} and in CeFeAs(O,F)\cite{CeFeAsOF_JZhao_01}, although the stacking order is not common.
This supports the importance of the FeAs intralayer coupling as well as  the weak interlayer coupling.

We now turn to the discussion of the relation between the structural distortion and the magnetic structure.
As mentioned before, the magnetic order in the FeAs system occurs either after the structural distortion or simultaneously.
A simple approach would be to consider a superexchange path between the nearest Fe neighbors.
Since the positional parameter of As and the lattice constant $c$ does not exhibit significant change, both the packing of the Fe-As layer and the Fe-As distance stay constant on passing through $T_0$.
Thus, the distortion within the Fe-As layer at {\tz} mainly results in a slight change in the Fe-Fe distance and the Fe-As-Fe bond angle.
The longer  $a$ lattice constant leads to a slightly smaller bond angle along $a$, $\angle$(Fe-As-Fe)$_a$=71.2$^{\circ}$ as compared to that along $b$,$\angle$(Fe-As-Fe)$_b$=72.1$^{\circ}$, in other words, the difference is less than 1$^{\circ}$.
It is unlikely that such minor distortions lead to significant differences in exchange parameters.
In contrast, it should be noted that the observed orthorhombic distortion and the columnar magnetic structure in {\sfa} are consistent with the results of a band structure calculation\cite{SrFe2As2_AJesche_01}.
The calculation gives both a comparable distortion and the correct magnetic structure in which the AFM arrangement is along the long $a$-axis. 
These calculation predict the distortion to be stable only for the columnar state, not for other magnetic structures or the non-magnetic state, indicating the strong coupling of magnetic order and orthorhombic distortion.
Similar results were also obtained for LaFeAsO.\cite{LaFeAsO_yildirim_01,LaFeAsOF_PVSushko_01}
For this compound, T. Yildirim discusses the lattice distortion to be related to a lifting of the double degeneracy of the columnar state within a localized spin model for a frustrated square lattice\cite{LaFeAsO_yildirim_01}.
However, I. I. Mazin \textit{et al.} questioned the applicability of such a model for these layered FeAs systems\cite{LaFeAsO_IIMazin_01}.
A detailed neutron scattering study on single crystals is expected to give further insights into this fascinating coupling.

The size of the magnetic moment of 1.01\,{\mub} determined by the present neutron diffraction study is the largest among the $R$FeAsO and the ternary arsenide $A$Fe$_2$As$_2$ reported so far: 0.36\,{\mub} for LaFeAs(O,F)\cite{LaFeAsOF_CdlCruz_01}, 0.8\,{\mub} for CeFeAs(O,F)\cite{CeFeAsOF_JZhao_01}, 0.87\,{\mub} for BaFe$_2$As$_2$\cite{BaFe2As2_QHuang_01}.
This corresponds to the stronger magnetism deduced for {\sfa} from bulk measurements, i.e. from the higher {\tz} and the larger hyperfine field in M\"ossbauer experiments.
On the other hand, it should be noted that the size of the ordered moment obtained in our neutron diffraction study is almost twice as large as that deduced in other microscopic measurements, ${\mu}$SR and M\"ossbauer spectroscopy.
Since the neutron diffraction intensity is proportional to the square of the size of the moment, this large difference can hardly be attributed to an experimental error.
Similar differences were obtained for {\bfa} and $R$FeAsO as well\cite{LaFeAsOF_HHKauss_01,BaFe2As2_MRotter_01}.
This points to a general problem.
While the neutrons directly probe the density of the magnetic moment, M\"ossbauer and ${\mu}$SR  rely on a scaling of the observed quantity, i.e. precession frequency and the hyperfine field.
It might be that these scaling procedures, which are well established for stable magnetic Fe systems with large moment, are not appropriate for the unusual magnetism in these layered FeAs systems.

\begin{acknowledgments}
We thank H.-H. Klaus and H. Rosner for stimulating discussions.
\end{acknowledgments}

\bibliography{kanekok,cond-mat}

\newpage

\newpage

\end{document}